\documentclass[%
 aip,
 amsmath,amssymb,
 reprint,%
]{revtex4-1}

\usepackage{graphicx}
\usepackage{dcolumn}
\usepackage{bm}

\usepackage[utf8]{inputenc}
\usepackage[T1]{fontenc}
\usepackage{mathptmx}

\newcommand{\BPCB} {(C$_5$H$_{12}$N)$_2$CuBr$_4$}
\newcommand{\DTN} {NiCl$_2$$\cdot4\,$SC(NH$_2$)$_2$}
\newcommand{\RbCuMoO}{Rb$_2$Cu$_2$Mo$_3$O$_{12}$}
\newcommand{\BaCdVOPO}{BaCdVO(PO$_4$)$_2$}

\begin{document}

\preprint{AIP/123-QED}

\title[]{Miniature capacitive Faraday force magnetometer for magnetization measurements at low temperatures and high magnetic fields}

\author{Dominic Blosser}
\affiliation{Laboratory for Solid State Physics, ETH Z\"urich, 8093 Z\"urich, Switzerland}
\author{Leonardo Facheris}
\affiliation{Laboratory for Solid State Physics, ETH Z\"urich, 8093 Z\"urich, Switzerland}
\author{Andrey Zheludev}
\affiliation{Laboratory for Solid State Physics, ETH Z\"urich, 8093 Z\"urich, Switzerland}

\date{\today}

\begin{abstract}
A Faraday force magnetometer is presented for measurements of magnetization at temperatures down to 100~mK and in magnetic fields up to 14~T. 
The specimen is mounted on a flexible cantilever forming a force-sensing capacitor in combination with a fixed back plate. 
Two different cantilever designs are presented. A torsion resistant cantilever allows to measure magnetization of highly anisotropic single crystal samples. 
Measurements of the metal organic quantum magnets \BPCB{} (BPCB) and \DTN{} (DTN) demonstrate the device's capabilities. 
Routinely, a specimen's magnetic moment is measured with a resolution better than $10^{-7}$~A$\,$m$^2$ ($10^{-4}$~emu). 
The device in miniaturized to fit is almost any cryostat. 
\end{abstract}

\maketitle

\section{\label{sec:intro}Introduction}

Magnetization is a very important property of materials. Consequently, a range of techniques have been developed allowing fast and reliable measurements of magnetization with excellent sensitivity and precision.
Most of these techniques are induction-based, requiring to mechanically move the sample through pick-up coils, inducing a time-varying magnetic flux. This includes the widely used techniques of vibrating sample magnetometry (VSM)\cite{Foner1959,Zieba1982,Foner1996} or  SQUID-based techniques\cite{Nagendran2011,Fagaly2006}.
At low temperatures, these methods become increasingly problematic as the mechanical motion causes unacceptable heating. For this reason, below a temperature of approx. 2~K, such methods cannot be used, and magnetization measurements become extremely challenging.

A Faraday force magnetometer works differently. In an external inhomogeneous magnetic field, a magnetic moment experiences a force pulling it towards regions where the field is strongest. This force which is directly proportional to magnetization is measured by a Faraday force magnetometer. Although apparently simple, this method is almost never used today as induction-based techniques provide much higher sensitivity and precision. 
At low temperatures, however, a Faraday force magnetometer has a crucial advantage: It does not require any mechanical motion and is essentially dissipation less. Therefore, Faraday force magnetometers are excellently suited for magnetization measurements at ultra-low temperatures. 
Although simple in principle, such measurements remain challenging in practice. Existing setups require rather specialized superconducting magnet systems and dedicated cryostats\cite{Sakakibara1994,Slobinsky2012}. 
A more fundamental challenge faced by these measurements is the potentially very large magnetic torque experienced by magnetized anisotropic samples in an external magnetic field. This requires carefully designed load cells that are insensitive to torque but still allow a highly sensitive detection of the magnetic force acting on the specimen\cite{Sakakibara1994}.

Here, we present a miniaturized and simple Faraday force magnetometer. 
The magnetic force is measured capacitively with the specimen mounted on a highly flexible cantilever. Different cantilever designs are presented that show excellent torsion resistance allowing to measure highly anisotropic samples.
Thanks to its compact design the magnetometer fits virtually any existing cryostat and superconducting magnet. In particular, it can be readily installed in the commercial Quantum Design dilution refrigerator with a sample space only 22~mm in diameter.

The paper proceeds as follows. In section \ref{sec:design} we describe the principle of the measurement and the design of the magnetometer. In section \ref{sec:performance} we present magnetization measurements of the nearly isotropic spin ladder material \BPCB{} (BPCB) and the extremely anisotropic compound \DTN{} (DTN) down to 100~mK and in fields up to 14~T. 
In the supplement we include the detailed specifications and technical drawings.

\section{\label{sec:design}Measurement procedure and Magnetometer design}

\subsection{Measurement procedure}
\begin{figure}
	\centering
	\includegraphics{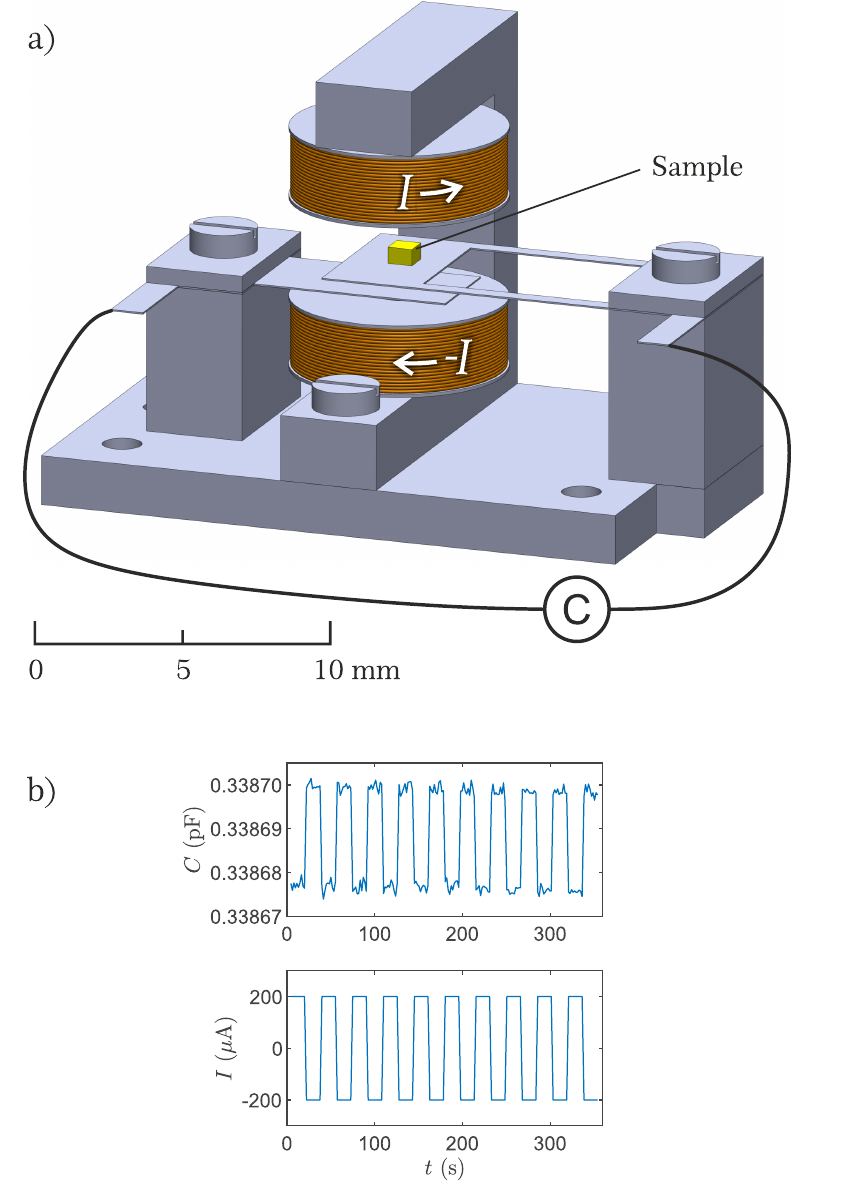}
	\caption{\label{fig:fmdr}
		a) Sketch of the Faraday force magnetometer. 
		b) Exemplary raw data showing the measured capacitance for a periodically reversed current in the gradient coils, i.e. a periodically reversed magnetic field gradient at the sample position.
	}
\end{figure}

In an external magnetic field $\mathbf{B}$, a magnetic moment $\mathbf{m}$ experiences both a force $\mathbf{F}=\left(\mathbf{m}\cdot\nabla\right)\mathbf{B}$ pulling it towards regions where the field is strongest, and a torque $\bm{\tau}=\mathbf{m}\times\mathbf{B}$ acting to align the magnetic moment with the field.
Applying a controlled magnetic field gradient and measuring the force $\mathbf{F}$, a sample's magnetization can be deduced. 
Implementing this idea, a device was built as sketched in Fig.~\ref{fig:fmdr}a). It is installed inside a cryostat and superconducting magnet system for measurements versus temperature and external magnetic field. 
A pair of small gradient coils generate a magnetic field gradient at the sample position. The sample is mounted on a flexible cantilever. Capacitance is measured between the cantilever and a fixed back-plate. This allows to deduce the deflection of the cantilever, thus the force on the sample and hence its magnetic moment.

We denote the symmetry axis of the gradient coils the $z$-axis. The cantilevers (see below) are only flexible in this direction. The measurement proceeds as follows: 
Opposite currents $\pm I$ are applied to the gradient coils generating no magnetic field but a  magnetic field gradient {$\frac{\partial B_z}{\partial z}\propto I$} at the sample position. The resulting magnetic force 
{$F_z \approx m_z \, \frac{\partial B_z}{\partial z}$} on the sample causes a small deflection $\delta\propto F_z$ of the cantilever.
Note that since $\nabla\cdot\mathbf{B}=0$, at the sample position there is also a radial field gradient. However, this leads to a radial force not significantly affecting the cantilever. 
Together the flexible cantilever and the fixed back plate approximately form a parallel plate capacitor with capacitance $C = \frac{ \epsilon_0 A}{d}$. Here $d$ denotes the spacing between the cantilever and the back-plate and $A$ is the area where they overlap.
Applying a small current $I$ to the gradient coils we measure a change in capacitance. Combining the above relations, we obtain
\begin{equation}
\left. \frac{dC}{dI} \right|_{I=0} \propto \left(\left.C\right|_{I=0}\right)^2 m_z.
\label{eq:FMDR}
\end{equation}
In practice, the quantity $\left. \frac{dC}{dI} \right|_{I=0}$ is measured by periodically reversing the current in the gradient coils and measuring capacitance continuously. Typical raw data from such a measurement is shown in Fig.~\ref{fig:fmdr}b).
Being able to \emph{actively control} the applied magnetic field gradient and to measure the \emph{derivative} $\left.\frac{dC}{dI} \right|_{I=0}$ is the crucial design ingredient. 
Any inhomogeneity of the externally applied static magnetic field will lead to an additional force acting on the sample. 
Furthermore, if the sample is not perfectly isotropic, magnetic torque can have an even bigger effect. 
However, by measuring the change of capacitance vs. a controlled magnetic field gradient, the measurement is unaffected by these contributions. The only requirement is that in the full measurement range the cantilever is only slightly deflected and the assumption of a parallel plate capacitor remains valid. 
The proportionality constant of Eq.~\ref{eq:FMDR} contains the geometry of the gradient coils, the elastic properties of the cantilever and the geometry of the capacitor formed by the cantilever and back plate. 
Since it is not possible to place different samples exactly at the same position on the cantilever, this Faraday force magnetometer setup cannot be reliably calibrated. Absolute units of magnetization are obtained by matching data taken at higher temperatures to measurements employing a standard inductive technique.

\subsection{Magnetometer}

The magnetometer (Fig.~\ref{fig:fmdr}a) is assembled on a Cu base plate small enough to fit in the 22~mm diameter sample space of a commercial Quantum Design dilution refrigerator\cite{QdDR}. The exchangeable cantilever and back plate are screwed onto two Cu mounts using brass screws. 
The Cu mounts are bonded to the base plate by a thin layer of epoxy ensuring they are electrically insulated. The height of the cantilever and back plate can be adjusted by inserting spacer plates to ensure the sample is positioned in the center of the gradient coils.

\begin{figure}
	\centering
	\includegraphics{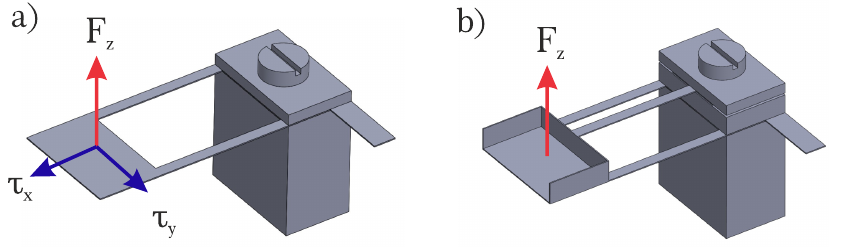}
	\caption{\label{fig:cantilever}
		a) Flat cantilever which is highly flexible to a vertical force $F_z$ but also strongly deformed by $\tau_x$ or $\tau_y$ torque components. 
		b) Three-leg-cantilever that is rigid to any torsion.}
\end{figure}

\subsection{Gradient coils}

The two gradient coils are made from superconducting NbTi wire wound on Stycast bodies with 750 turns each. They are mounted on a Stycast support in an approximate Helmholtz geometry. 
Due to their small size and the small spacing between the coils, even a low current of $I=30$~mA creates a sizable field gradient of 2.3~T/m at the sample position.
Below a temperature of 1~K, these coils remain superconducting up to 14~T. The upper critical field is reduced to 13~T at 2~K and 10~T at 4~K, respectively.

\subsection{Cantilever}

First and foremost the cantilever needs to be highly flexible in the vertical direction.
Fig.~\ref{fig:cantilever} shows two different cantilever designs.
A simple flat cantilever (Fig.~\ref{fig:cantilever}a) is excellently suited for nearly isotropic samples. It is extremely flexible. Thanks to its two legs it shows some rigidity with respect to torsion along $x$. However, any flat cantilever will show a similar response to a torque $\tau_y$ as to a force $F_z$. Therefore, neither this nor any other flat cantilever is suitable for anisotropic samples subject to magnetic torque. 
A three-leg-cantilever (Fig.~\ref{fig:cantilever}b) is rigid to any torsion and still very soft to a vertical force $F_z$.  
Both cantilevers are made from $50\,\mu$m CuBe foil.

\begin{figure}
	\centering
	\includegraphics{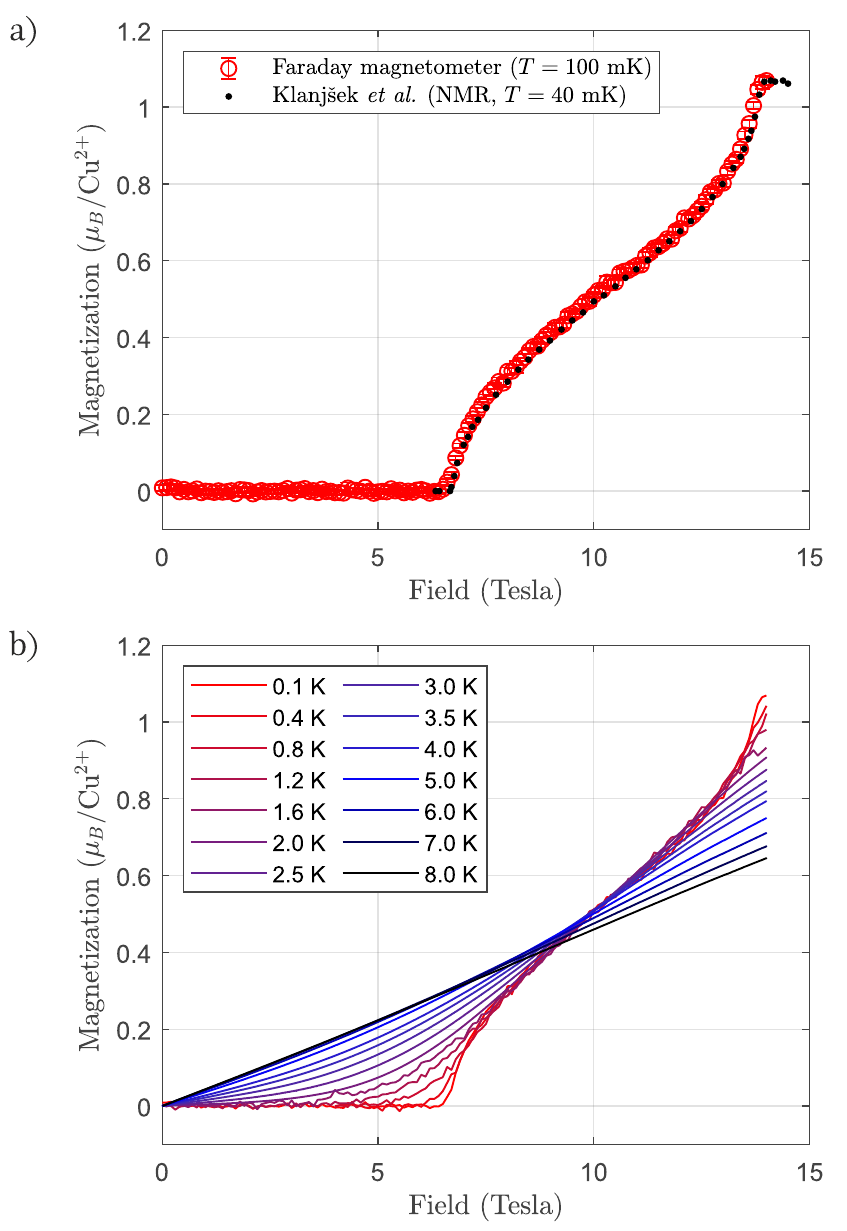}
	\caption{\label{fig:bpcb}
a) Magnetization of \BPCB{} (BPCB) measured on a 0.69(7)~mg single crystal at 100~mK using the flat cantilever. 
b) Magnetization at various temperatures. Data above 2~K are measured using a commercial vibrating sample magnetometer.
}
\end{figure}

\begin{figure}
	\centering
	\includegraphics{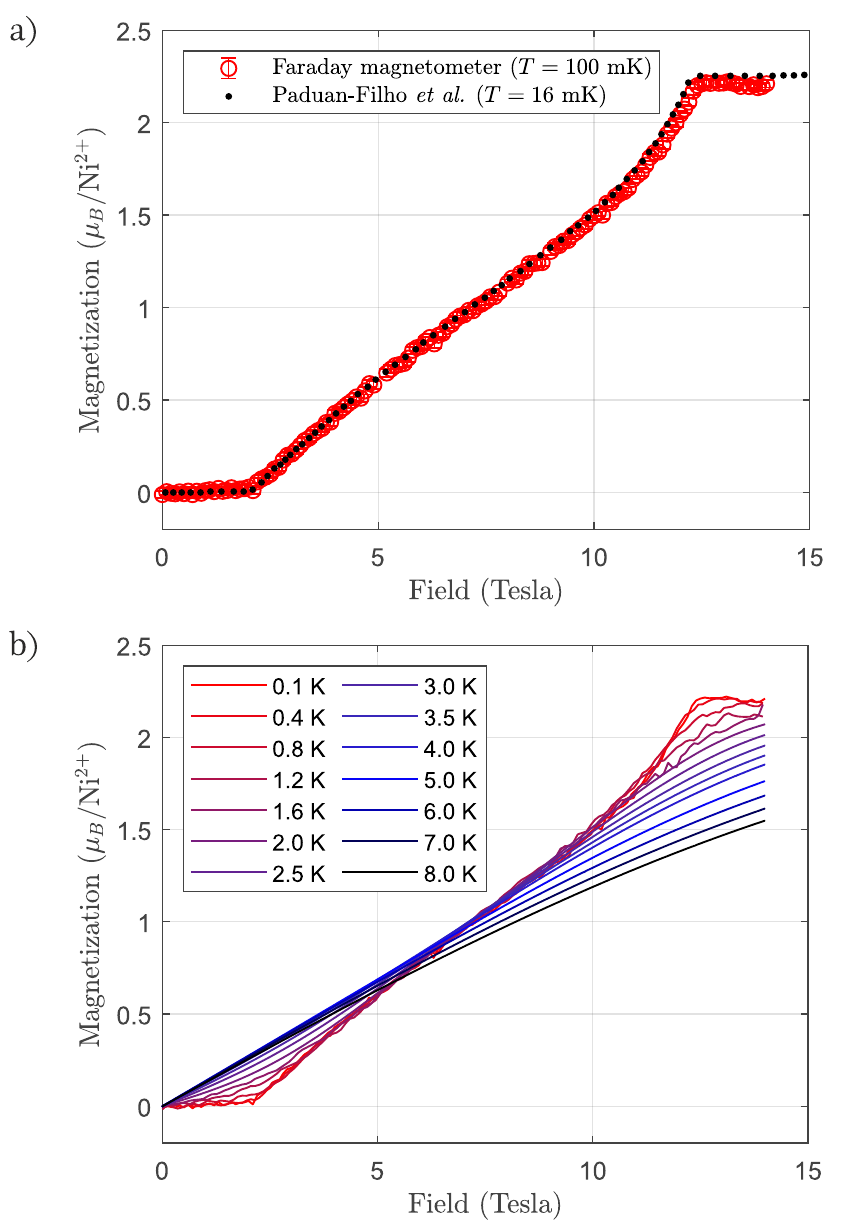}
	\caption{\label{fig:dtn}
		a) Magnetization of \DTN{} (DTN) measured on a 0.7(1)~mg single crystal at 100~mK using the three-leg-cantilever. 
b) Magnetization at various temperatures. Data above 2~K are measured using a commercial vibrating sample magnetometer.
}
\end{figure}

\subsection{Capacitance measurement}
The magnetometer presented here requires very precise measurements of capacitance. The  change in capacitance upon applying a small field gradient can be very small and a precision of $10^{-6}$~pF is often required. 
Indeed, such precision is achievable using commercial instruments\cite{AHCapacitanceBridge}. However, fully shielded coaxial cables connecting the cantilever and back plate to the capacitance bridge are absolutely required.

\section{\label{sec:performance}Performance and sensitivity}

\subsection{Nearly isotropic $S=1/2$ spin ladder compound BPCB}

Magnetization of the nearly isotropic $S=1/2$ spin ladder compound \BPCB{} (BPCB) was measured. In this compound the $S=1/2$ spins localized on the Cu$^{2+}$ ions interact antiferromagnetically to form weakly coupled dimers\cite{Patyal1990}. In the absence of a magnetic field these show a non-magnetic singlet groundstate. In an applied magnetic field of $B_{c1}=6.66(6)$~T the system becomes partially magnetized before it fully saturates at $B_{c2}=13.6(1)$~T\cite{ThielemannRuegg2009,Blosser2018}. 

A 0.69(7)~mg ($\sim1\times0.5\times0.5$ mm) BPCB single crystal sample was fixed on the flat cantilever with a small amount of Apiezon N grease with the crystallographic $b$ direction parallel to the external magnetic field. Data obtained at 2, 3 and 4~K show perfect agreement with data taken on a bigger sample using a commercial Quantum Design vibrating sample magnetometer for the PPMS\cite{QdVSM} after fixing an overall prefactor. 
Magnetization measured at 100~mK is shown in Fig.~\ref{fig:bpcb}a). It shows good agreement with nuclear magnetic resonance (NMR) data of Ref.~\onlinecite{Klanjsek2008}. The very small discrepancy is likely due to the slightly different orientation of the magnetic field in the NMR experiments.  
Magnetization at different temperatures is shown in Fig.~\ref{fig:bpcb}b). Data above 2~K are measured using the vibrating sample magnetometer. 

Capacitance was measured using an Andeen-Hagerling capacitance bridge\cite{AHCapacitanceBridge} with a high integration time providing one reading of capacitance roughly every 3~s.
We have applied a current of $\pm200$~$\mu$A to the gradient coils. In total, for each point 200 readings of capacitance were recorded while reversing the applied field gradient 10~times (c.f. Fig.~\ref{fig:fmdr}b). One measurement of magnetization at a fixed temperature and magnetic field value hence took approx. 12 minutes. 
The overall change of capacitance due to magnetic torque and inhomogeneity of the external magnetic field was less than 2\% throughout the whole measurement range. Thus the assumption of a parallel plate capacitor formed by the cantilever and back-plate is well justified and the flat cantilever can safely be used.

\subsection{Anisotropic $S=1$ magnet DTN}

As a second test case, we have measured magnetization of \DTN{} (DTN). In this compound the Ni$^{2+}$ ions carry an $S=1$ spin and strong easy plane single ion anisotropy favours the $S_z=0$ state. In the absence of a magnetic field this leads to a gapped non-magnetic groundstate. 
Applying a magnetic field along the anisotropy direction this spin gap is closed at a critical field of $H_{c1}=2.11(1)$~T. Beyond this field the system is partially magnetized and shows long-range magnetic order. At a field of $H_{c2}=12.11(3)$~T it becomes fully polarized\cite{PaduanFilho2004}. 

In an applied magnetic field, this highly anisotropic compound experiences strong magnetic torque. Mounted on the flat cantilever, even for a very small sample, this leads to a severe bending of the cantilever and eventually to a mechanical touch of the cantilever and back-plate. Using a $0.7(1)$~mg single crystal mounted on the three-leg-cantilever, however, no such issues are encountered and the overall change of capacitance is again less than 2\% throughout the whole measurement range.
Similar measurement parameters were used as for BPCB, but with a current of $\pm400$~$\mu$A applied to the gradient coils. Absolute units were obtained by matching data obtained at 2, 3, and 4~K to VSM measurements measured on a bigger sample. 

Magnetization measured at 100~mK is shown in Fig.~\ref{fig:dtn}a). It agrees well with measurements by A. Paduan-Filho {\it et al.}\cite{PaduanFilho2004} (at 16~mK), except for a small overall discrepancy of $\approx 2$\% of the saturated moment. Most likely, this slight mismatch is due to uncertainty in the sample mass. When normalized to magnetization at saturation the agreement is excellent. 
Fig.~\ref{fig:dtn}b) shows magnetization measured at different temperatures. Data above 2~K are measured on a bigger sample using the vibrating sample magnetometer.

\subsection{Background}

Measurements of the empty CuBe cantilevers revealed weakly diamagnetic behavior on top of a small paramagnetic contribution, in total resembling a slanted Brillouin function. This background was subtracted from all measurements. It never exceeded 1\% of the sample's saturation magnetization for the flat cantilever or 3\% for the three-leg-cantilever, respectively.

\subsection{Sensitivity}

As a measure for sensitivity, we use the standard deviation of all points where magnetization is zero (below 6.6~T for BPCB, or 2.1~T for DTN, respectively). This gives 3$\cdot10^{-5}$~emu, or 1$\cdot10^{-4}$~emu, respectively, for the two measurements. In both cases this corresponds to less than 0.5\% of the saturated moment of these small ($\approx 0.7$~mg) samples. 

The sensitivity is determined by (i) the precision of the capacitance measurements, (ii) the overall capacitance and (iii) the magnitude of the applied magnetic field gradient. 
i) Using a state-of-the-art capacitance bridge and fully shielded measurement cables, very precise capacitance measurements are possible. The precision of these measurements can only be improved easily by averaging repeated measurements.
ii) The overall capacitance 
could be increased by using larger capacitor plates or by bringing them closer together. At the same time, however, any twisting of the cantilever due to magnetic torque will then lead to much more severe deviations from the parallel plate geometry and the potential for improved sensitivity is limited. 
iii) Applying a higher current to the gradient coils (creating a larger field gradient at the sample position) directly leads to a larger signal. In our setup the gradient coils are superconducting but the wires supplying the current to the coils are not. This leads to heating at low temperatures and effectively limits the maximum current to a few 100 $\mu$A at 100~mK. Using exclusively superconducting wires this current could easily be increased 10 or even 100-fold. This would directly improve sensitivity by the same amount.

%

\section{\label{sec:conclusion}Conclusion}

A capacitive Faraday force magnetometer for magnetization measurements at temperatures down to 100~mK and in magnetic fields up to 14~T was presented. Two highly flexible force-sensing cantilevers were described.
Magnetization measurements of the nearly isotropic compound \BPCB{} (BPCB) and the highly anisotropic magnet \DTN{} (DTN) demonstrate the device's performance. In both cases a $\approx0.7$~mg single crystal sample was measured and the resolution was better than 0.5\% of the saturated moment.  
The presented Faraday force magnetometer is miniaturized ($2\times2\times2$~cm) to fit into almost any cryostat.


\section*{Further measurements}
The presented data on BPCB are discussed in more detail in Ref.~\onlinecite{Blosser2019Thesis}. In addition, this device was employed for magnetization measurements on the compounds \RbCuMoO{} (Ref.~\onlinecite{Hayashida2019}) and \BaCdVOPO{} (Ref.~\onlinecite{Bhartiya2019})

\begin{acknowledgments}
We thank Urs Notter, Willy Staubli and Andreas Stuker from the mechanical workshop at ETH Z\"urich for their help with the manufacturing of the magnetometer.
This work was partially supported by the Swiss National Science Foundation under Division II. 
\end{acknowledgments}


%
\section*{References}
\bibliography{BibDB2,BibDB_InstrManuals,BibDB_MyPublications}

%
%

\begin{widetext}

\clearpage\newpage
\noindent
\makebox[\textwidth]{
	\includegraphics[page=1]{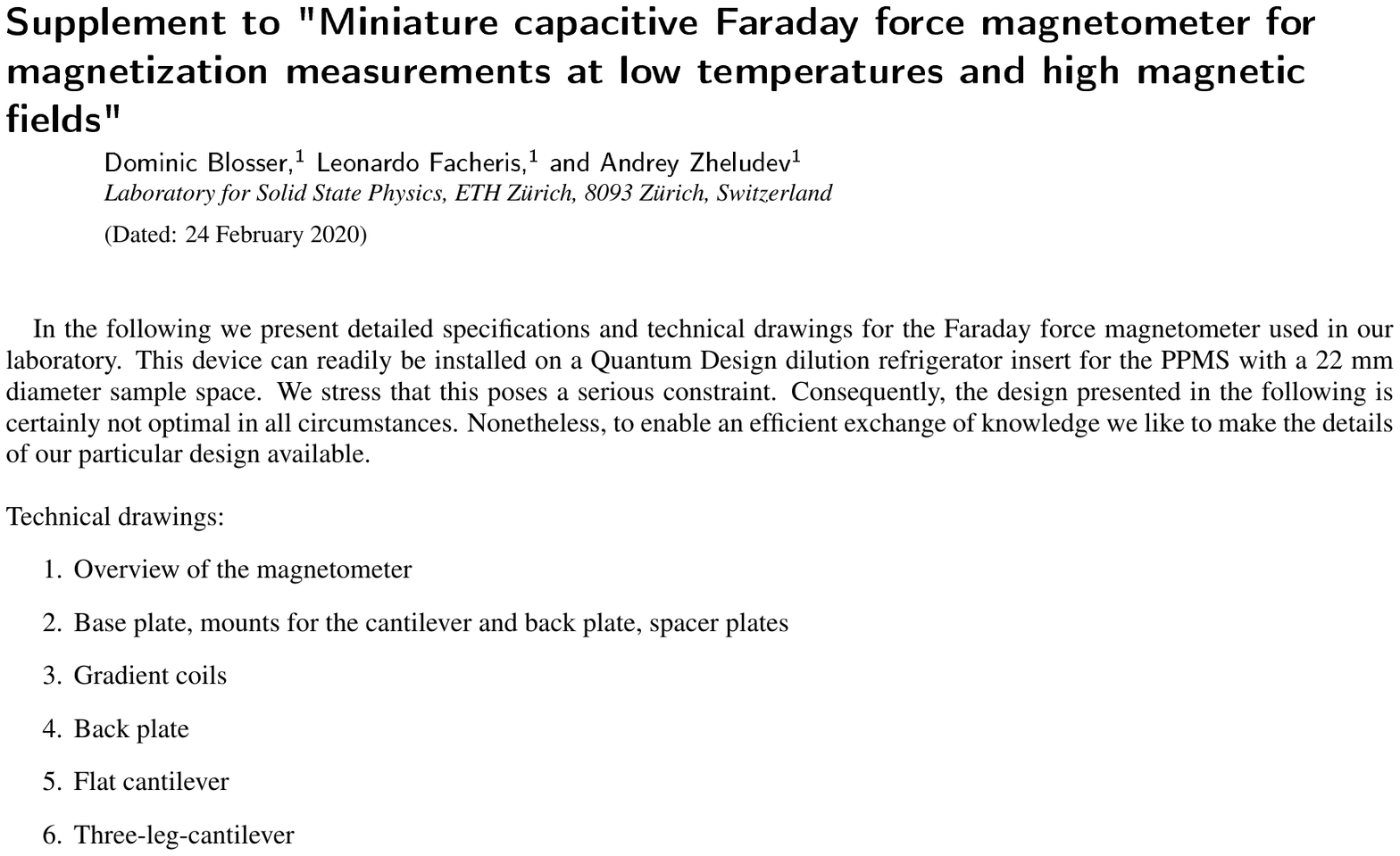}
}

\clearpage\clearpage\newpage
\noindent
\makebox[\textwidth]{
	\includegraphics[trim={0 0 0 1.3cm},clip,page=1,scale=0.9]{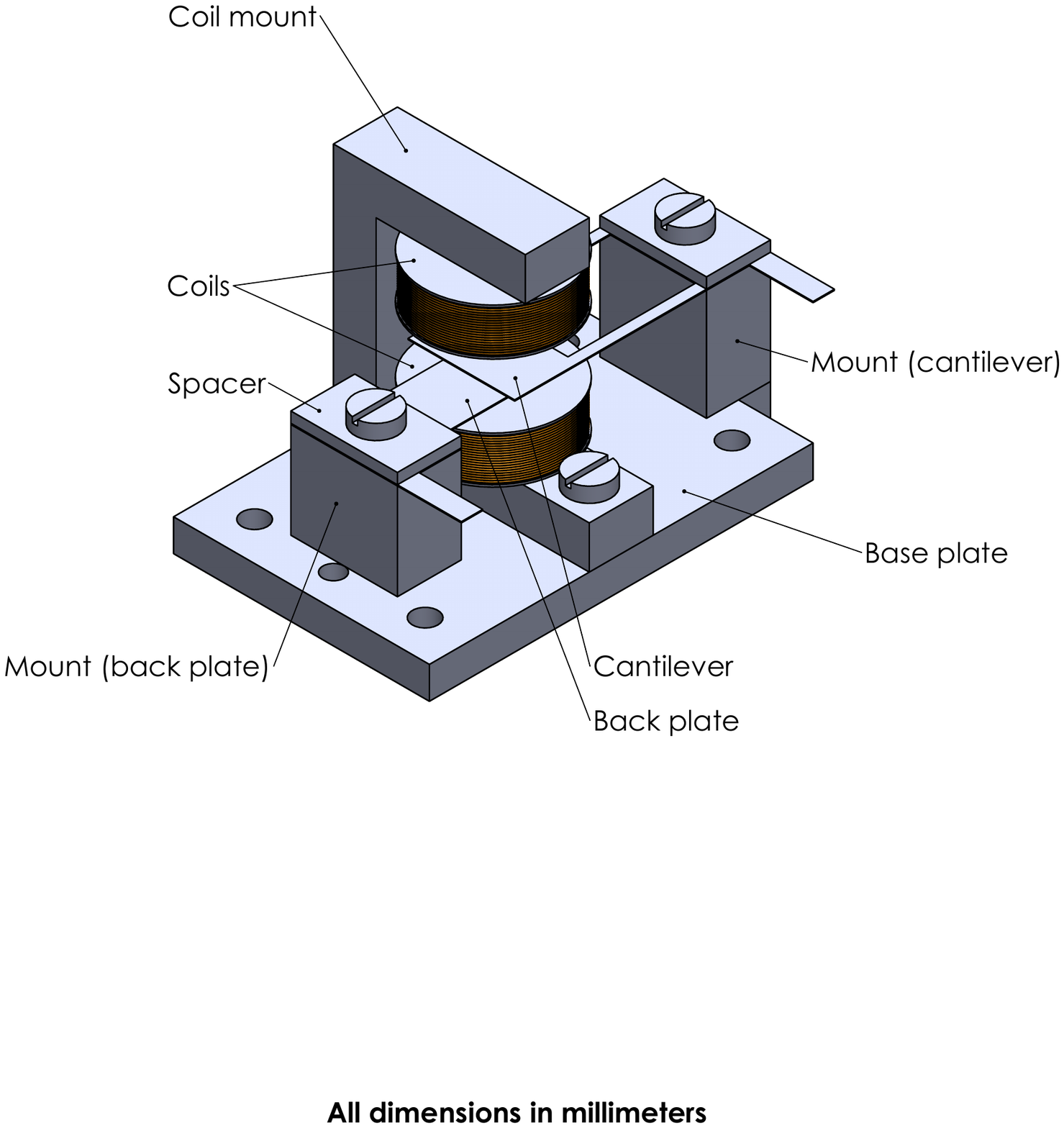}
}

\clearpage\clearpage\newpage
\noindent
\makebox[\textwidth]{
	\includegraphics[trim={0 0 0 1.3cm},clip,page=2,scale=0.9]{Drawings.PDF}
}

\clearpage\clearpage\newpage
\noindent
\makebox[\textwidth]{
	\includegraphics[trim={0 0 0 1.3cm},clip,page=3,scale=0.9]{Drawings.PDF}
}

\clearpage\clearpage\newpage
\noindent
\makebox[\textwidth]{
	\includegraphics[trim={0 0 0 1.3cm},clip,page=4,scale=0.9]{Drawings.PDF}
}

\clearpage\clearpage\newpage
\noindent
\makebox[\textwidth]{
	\includegraphics[trim={0 0 0 1.3cm},clip,page=5,scale=0.9]{Drawings.PDF}
}

\clearpage\clearpage\newpage
\noindent
\makebox[\textwidth]{
	\includegraphics[trim={0 0 0 1.3cm},clip,page=6,scale=0.9]{Drawings.PDF}
}

\end{widetext}
\end{document}